\documentclass[twocolumn,showpacs,preprintnumbers,amsmath,amssymb]{revtex4}

\usepackage{graphicx}
\usepackage{dcolumn}
\usepackage{bm}

\begin{document}

\preprint{NEWwigner080214}

\title{Quantum Correlation Bounds for Quantum Information Experiments Optimization:
the Wigner Inequality Case }

\author{F. A. Bovino }%
 \email{Fabio.Bovino@elsagdatamat.com}

\affiliation{%
Elsag Datamat  \\
Via Puccini 2-16154 Genova (Italy)
}%

\author{I. P. Degiovanni}%
 \email{i.degiovanni@inrim.it}

\affiliation{
Istituto Nazionale di Ricerca Metrologica \\
Strada delle Cacce 91-10135 Torino (Italy)
}%

\date{\today}

\begin{abstract}
Violation of modified Wigner inequality by means binary bipartite
quantum system  allows the discrimination between the quantum world
and the classical local-realistic one, and also ensures the security
of Ekert-like quantum key distribution protocol. In this paper we
study both theoretically and experimentally the bounds of quantum
correlation associated to the modified Wigner's inequality finding
the optimal experimental configuration for its maximal violation. We
also extend this analysis to the implementation of Ekert's protocol.
\end{abstract}

\maketitle

The famous EPR paper \cite{epr} originated the debate on the
existence of local-realistic hidden variables theory able to replace
quantum theory. The Bohm's version of EPR argument \cite{bohm}
dealing with the quantum correlation of two-particle entangled
state, triggered Bell's derivation of an experimentally testable
inequality \cite{bell}, in principle allowing the discrimination
between the quantum world and the classical local-realistic one.

Since then, several Bell's inequalities using two or more particles
have been proposed \cite{wigner70,otine,GENrev}, and a lot of
experiments with different quantum systems have been performed
showing violation of Bell's inequality in good agreement with the
predictions of quantum mechanics \cite{GENrev,bellexp,zeilingerEXP}.
Furthermore the effect of practical inefficiencies has been studied
\cite{BOVLAM}, even if an experiment "ruling out" all the loopholes
together has not yet been realized \cite{GENrev}.

More recently, several studies on the \textit{extension} of quantum
correlation appeared, such as studies on the possibility of
super-quantum correlation \cite{POP94}, on the relative extension of
quantum correlations versus the classical ones \cite{CAB05}, on the
bounds of quantum correlation associated to
Clauser-Horne-Shimony-Holt (CHSH) inequality \cite{CAB04, FS04,
BCC04}.

On the other side, from the early days of quantum information it is
clear that quantum correlation (entanglement), and Bell's inequality
as mean to highlight its presence, has a central role in this new
field. In this context, the most advanced application in quantum
information is quantum key distribution (QKD)
\cite{BB84,gisinrevmod}, various systems of QKD have been
implemented and tested by groups around the world
\cite{gisinrevmod}. In 1991 A. Ekert proposed a new QKD protocol
whose security relies on the non-local behavior of quantum
mechanics, i.e., on Bell's inequalities \cite{EKE91}. The firsts
experimental implementations of Ekert's protocol were performed nine
years later by Naik \textit{et al.} \cite{qk3} implementing a
variant of this protocol based on CHSH inequality, and by Jennewein
\textit{et al.} \cite{qk2} implementing a variant of this protocol
based on the Wigner's inequality (WI).

In ref. \cite{qk2} the violation of WI exploiting a two-photon
singlet state was first proposed to provide an easier and equally
reliable eavesdropping test. Unfortunately, this is not the case. We
proved both theoretically and experimentally that if the
eavesdropper has the control of both channels, he is able to violate
WI with separable states \cite{CDR03,BCC03}. Furthermore we proposed
a \textit{modified} version of WI whose violation ensures the
security of the communication \cite{CDR03, BCC03}. (Recently was
brought to our attention Ref. \cite{Koc93}, where the algebraic
properties of the singlet state in connection with the original
Wigner's inequality are investigated, arguing that the original
Wigner's argument is not significant in deriving conclusions about
local realism. It is noteworthy to observe that in the derivation of
the modified WI there is not any assumption on the quantum state
considered \cite{CDR03}, thus Koc's considerations cannot be applied
in the case of the modified WI).

Scope of this paper is to study theoretically and experimentally the
bounds of quantum correlation associated to the modified WI, aiming
not only to investigate the existence of super-quantum correlation,
but mainly to find configuration for the maximal possible violation
of modified WI, and for the optimal implementation of Ekert's QKD
protocol based on this inequality.

\begin{figure}[tbp]
\par
\begin{center}
\includegraphics[angle=0, width=10 cm]{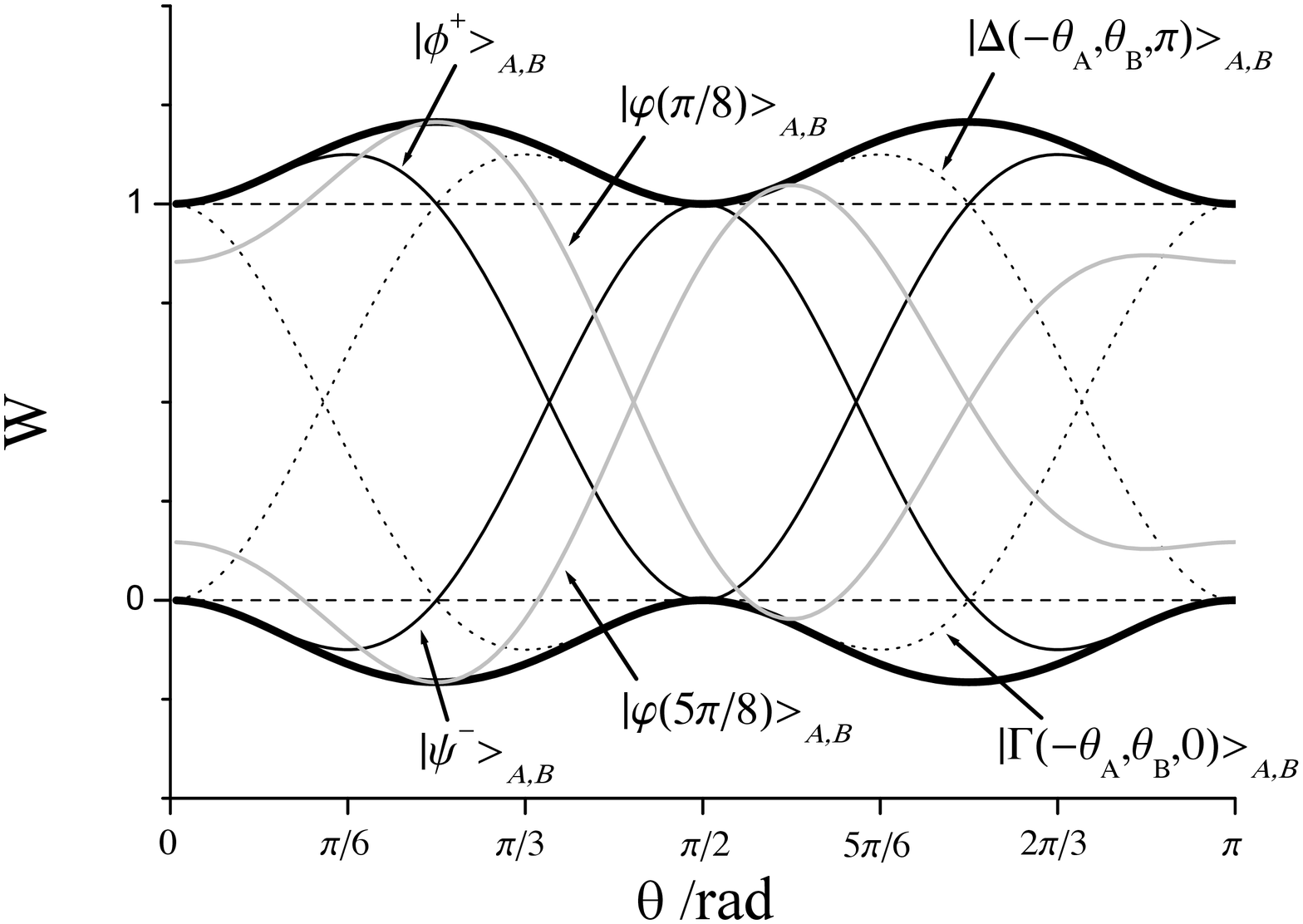}
\end{center}
\caption{ Theoretical curves. Wigner's parameter classical
correlation bounds are $ W=0$ and $ W= 1$. Thick black lines are the
minimum and maximum eigenvalues of the operator $\widehat{W}$ (upper
and lower bounds of quantum correlation). The maximal upper (lower)
violation of WI can be schieved e.g. with $| \varphi(\pi/8)
\rangle_{A,B}$ ($| \varphi(5 \pi/8) \rangle_{A,B}$) (light gray
lines). Curves corresponding to W obtained averaging on $| \psi^{-}
\rangle_{A,B}$ and $| \phi^{+} \rangle_{A,B}$ (thin black line) and
on $|\Gamma(-\theta_{A}, \theta_{B},0)\rangle_{A,B}$ and
$|\Delta(-\theta_{A}, \theta_{B},\pi)\rangle_{A,B}$ (dotted black
line) are also shown.} \label{Figure 1}
\end{figure}

We consider the operator $\widehat{W}$ associated to the WI with a
parametrization similar to the one proposed by Filipp and Svozil for
the CHSH inequality \cite{CAB04, FS04, BCC04}
\begin{eqnarray}\label{Wop}
\widehat{W}&=&\widehat{P}_{+,A}(-\theta)\otimes
\widehat{P}_{+,B}(0)+\widehat{P}_{+,A}(0)\otimes
\widehat{P}_{+,B}(\theta) \nonumber \\
& & +\widehat{P}_{-,A}(0)\otimes \widehat{P}_{-,B}(0)-
\widehat{P}_{+, A}(-\theta)\otimes \widehat{P}_{+ ,B}(\theta),
\label{Wop}
\end{eqnarray}
where $\widehat{P}_{+,i}(\theta)$ is the projector on the state $|
s_{+}(\theta) \rangle_{i}=\cos(\theta)| H \rangle_{i}+\sin(\theta)|
V \rangle_{i}$ of the $i$-th subsystem (with $i=A,B$), and
$\widehat{P}_{-,i}(\theta)$ is the projector on the orthogonal state
$| s_{-}(\theta) \rangle_{i}=\cos(\theta)| V
\rangle_{i}-\sin(\theta)| H \rangle_{i}$.

Thus, the modified Wigner's parameter defined in \cite{CDR03,BCC03}
is obtained as $W=\mathrm{Tr}(\widehat{W} ~ \widehat{\rho})$, with
$\widehat{\rho}$ being the density matrix of our quantum system.
Thus, we obtain the expression for $W$ in Filipp-Svozil
parametrization analogous to the one of \cite{CDR03, BCC03}
\begin{eqnarray}
W& =&
p_{-\theta_{A},0_{B}}(+_{A},+_{B})+p_{0_{A},\theta_{B}}(+_{A},+
_{B})  \nonumber \\
& &+ p_{0_{A},0_{B,}}(-_{A},-_{B})
-p_{-\theta_{A},\theta_{B,}}(+_{A},+_{B}),    \label{Wnum}
\end{eqnarray}
with
$p_{\alpha_{A},\beta_{B}}(\pm_{A},\pm_{B})=\mathrm{Tr}\left[\widehat{P}_{\pm,A}(\alpha)\otimes
\widehat{P}_{\pm,B}(\beta) ~ \widehat{\rho}\right]$.

As discussed in \cite{CDR03,BCC03} the classical limit for the
modified WI is $W\geq 0$, while the maximum violation obtainable
with the singlet state
$|\psi^{(-)}\rangle_{A,B}=2^{-1/2}(|H\rangle_{A} | V\rangle_{B}
-|V\rangle_{A} | H\rangle_{B})$  is $W= -0.125$ obtained when
$\theta=\pi/6$. We also showed that the violation of $W\geq 0$
guarantees the security of the Ekert's protocol based on the
modified WI.

To study the bounds of quantum correlation for the modified WI we
exploited the $min/max$ principle. In Fig. 1 we plot the minimum and
the maximum eigenvalues of the operator $\widehat{W}$ in Eq.
(\ref{Wop}) (black thick lines), while the other two eigenvalues do
not violate the bound of classical correlation. The first
observation is that $\widehat{W}$ presents both an upper and lower
bounds, and that the singlet state is not the one producing the
maximal violation. In fact the extremal values for the modified WI
are $ -0.20711 $ and $ 1.20711 $ obtained for e.g. $ \theta= \pi/4
$.

The presence of the quantum correlation upper bound for WI leaded us
to investigate on the existence of an upper bound also for classical
correlation. With this aim we reconsider the Wigner's argument
\cite{wigner70, CDR03}. The assumptions on locality and realism in
the derivation of the (modified) WI are embedded in the classical
probability distribution, $\mathcal{P}(x_{1},x_{2};y_{2},y_{3})$,
where $x_{1}$ and $x_{2}$ are the hidden variables associated with
the physical property inducing Alice's outcomes associated to the
projection on $|s_{x_{1}}(-\theta)\rangle_{A}$ and
$|s_{x_{2}}(0)\rangle_{A}$ respectively. Analogously $y_{2}$ and
$y_{3}$ correspond to the physical property inducing Bob's outcomes
associated to the projection on $|s_{y_{2}}(0)\rangle_{B}$ and
$|s_{y_{3}}(\theta)\rangle_{B}$. Thus, we identify the possible
values of $x_{1,2}$ and $y_{2,3}$ with Alice and Bob's measurement
outcomes, in other words $x_{1,2}=+_{A},-_{A}$ and
$y_{2,3}=+_{B},-_{B}$. Following this approach we write,
\begin{eqnarray}
p_{-\theta _{A},0 _{B}}(+_{A},+_{B}) &=&\sum_{x_{2},y_{3}}%
\mathcal{P}(+_{A},x_{2};+_{B},y_{3}), \nonumber\\
p_{0 _{A},\theta _{B}}(+_{A},+_{B}) &=&\sum_{x_{1},y_{2}}%
\mathcal{P}(x_{1}, +_{A};y_{2},+_{B}), \nonumber \\
p_{-\theta _{A},\theta _{B}}(+_{A},+_{B}) &=&\sum_{x_{2},y_{2}}%
\mathcal{P}( +_{A},x_{2};y_{2},+_{B}), \nonumber \\
p_{0 _{A},0 _{B}}(-_{A},-_{B}) &=&\sum_{x_{1},y_{3}}%
\mathcal{P}( x_{1}, -_{A};-_{B},y_{2}) . \label{Pcal}
\end{eqnarray}

Substituting Eq.s (\ref{Pcal}) in Eq. (\ref{Wnum}) we obtain
\begin{eqnarray}
W&=&\mathcal{P}(+_{A},+_{A};+_{B},+_{B})+\mathcal{P}(+_{A},+_{A};+_{B},-_{B})
\nonumber
\\ & &+
\mathcal{P}(-_{A},+_{A};+_{B},+_{B})+\mathcal{P}(+_{A},-_{A};+_{B},-_{B})
\nonumber \\ & &
+\mathcal{P}(-_{A},+_{A};-_{B},+_{B})+\mathcal{P}(+_{A},-_{A};-_{B},-_{B})
\nonumber
\\ & &+ \mathcal{P}(-_{A},-_{A};-_{B},+_{B})
+\mathcal{P}(-_{A},-_{A};-_{B},-_{B})\leq 1 \nonumber
\end{eqnarray}
where the last inequality is obtained by exploiting the
normalization condition $\sum_{x_{1}, x_{2},y_{2},y_{3}}
\mathcal{P}(x_{1}, x_{2},y_{2},y_{3})=1$. This is to our knowledge
the first derivation of the upper bound of classical correlation
associated to the (modified) WI.

Summarizing the Wigner's parameter bounds associated to
local-realistic theory are $0 \leq W \leq 1$, while the bounds of
quantum correlations in Filipp-Svozil parametrization are $-0.20711
\leq W \leq 1.20711$ at specific choices of the angle $\theta$.
Furthermore the states producing the maximal violation of the
inequality $0 \leq W \leq 1$ for the upper (lower) bound is given by
the eigenstate corresponding to the maximum (minimum) eigenvalue of
$\widehat{W}$, and in none of these two cases it corresponds to the
singlet state used in previous experiment \cite{zeilingerEXP}. In
particular, the eigenstates corresponding to the maximum and the
minimum eigenvalues of modified WI are of the form
\begin{eqnarray}\nonumber
| \varphi(\xi) \rangle_{A,B} &=& \cos(\xi) | \phi^{+}
\rangle_{A,B}+\sin(\xi) | \psi^{-} \rangle_{A,B}=\\
&=&\frac{1}{\sqrt{2}}(
|H\rangle_{A}|s_{+}(\xi)\rangle_{B}-|V\rangle_{A}|s_{-}(\xi)\rangle_{B})
\label{state}
\end{eqnarray}
with $|\phi^{(+)}\rangle_{A,B}=2^{-1/2}(|H\rangle_{A} | H\rangle_{B}
+|V\rangle_{A} | V\rangle_{B})$. For example, in the case of
$\theta=\pi/4$ the maximal violation of the upper bound is obtained
when $\xi=\pi/8$ (upper light gray line in Fig. 1), while the
maximal violation of lower bound is obtained when $\xi=5 \pi/8$
(lower light gray line in Fig.1).

\begin{figure}[tbp]
\par
\begin{center}
\includegraphics[angle=0, width=8 cm]{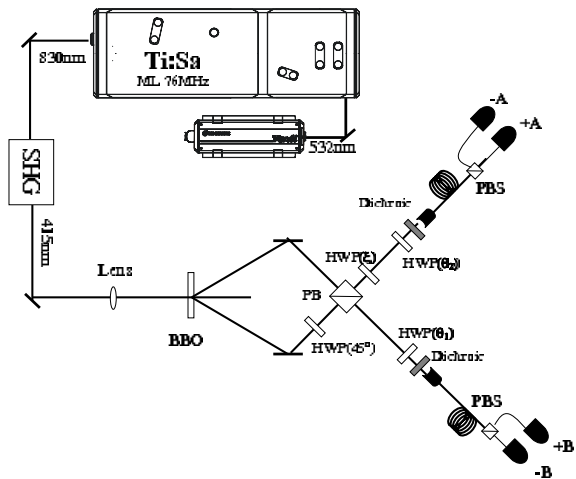}
\end{center}
\caption{Experimental setup. } \label{Figure setup}
\end{figure}

According to the theoretical arguments discussed above it is of
straightforward interest to perform an experiment to test the
behavior of the quantum correlations for the  modified WI. We
perform this experiment exploiting the optical setup presented in
Fig. 2, and analogous to the one we used in Ref. \cite{BCC04}. The
source of pulsed parametric down-conversion (PDC) is obtained by a 3
mm length BBO nonlinear crystal pumped with ultrashort pump pulses
(160 fs) at 415 nm generated from the second harmonic of a
mode-locked Ti-Sapphire with repetition rate of 76 MHz. PDC
degenerate photon pairs at 830 nm are generated by a non-collinear
type II phase matching, providing eventually a polarization
entanglement, i.e. the singlet state $| \psi^{-} \rangle_{A,B}$
\cite{kwiat2}, when time-compensated PDC scheme is applied
\cite{kimgrice}. To realize the set of entangled states $| \varphi
(\xi) \rangle_{A,B}$ in Eq. (\ref{state}) an half-wave plate (HWP)
on the channel $B$ is used to rotate the polarization of photon $B$.

The local measurements on photon $A$ and $B$ are performed by
identical apparatuses composed of open-air fiber couplers collecting
the PDC in single-mode optical fibers. HWPs before the fiber coupler
together with fiber-integrated polarizing beam splitters (PBSs) and
fiber polarization controllers project photons in the polarization
basis. Photons at the output ports of the PBSs are detected by fiber
coupled commercial photon counters. Dichroic mirror are placed in
front of the fiber couplers to reduce stray-light. Coincident counts
are measured by a non-commercial prototype of four-channel
coincident circuit. Single-counts and coincidences are counted by a
sixteen channels counter plug-in PC card.

Fig. 3 shows highly stable and repeatable $W$ measurement points
(dots) plotted versus $\theta$, and the various curves are
associated to different values of the parameter $\xi$. The thicker
curves correspond to the theoretically predicted $W$ quantum
correlation bounds. There is a good qualitative and quantitative
agreement between theoretical and experimental bounds, even if the
experimental upper (lower) bounds stands slightly below (above) the
theoretical predictions. These effects are, as usual, imputable to
noise and imperfections associated to the polarization preservation
and measurement of some setup components, namely HWPs, PBSs, and
fibers. In fact, the discrepancy between the theoretical and the
experimental results observed is confirmed by an equivalent noise
level as verified during the alignment process, e.g. at specific
angle settings of the polarizers \cite{BOVLAM}. Fig. 4 shows the
contour plot corresponding to the experimental data in Fig. 3. This
plot highlights the region of values of $\xi$ and $\theta$ where we
observed violations of both upper and lower bounds of the modified
WI.

\begin{figure}[tbp]
\par
\begin{center}
\includegraphics[angle=0, width=10 cm]{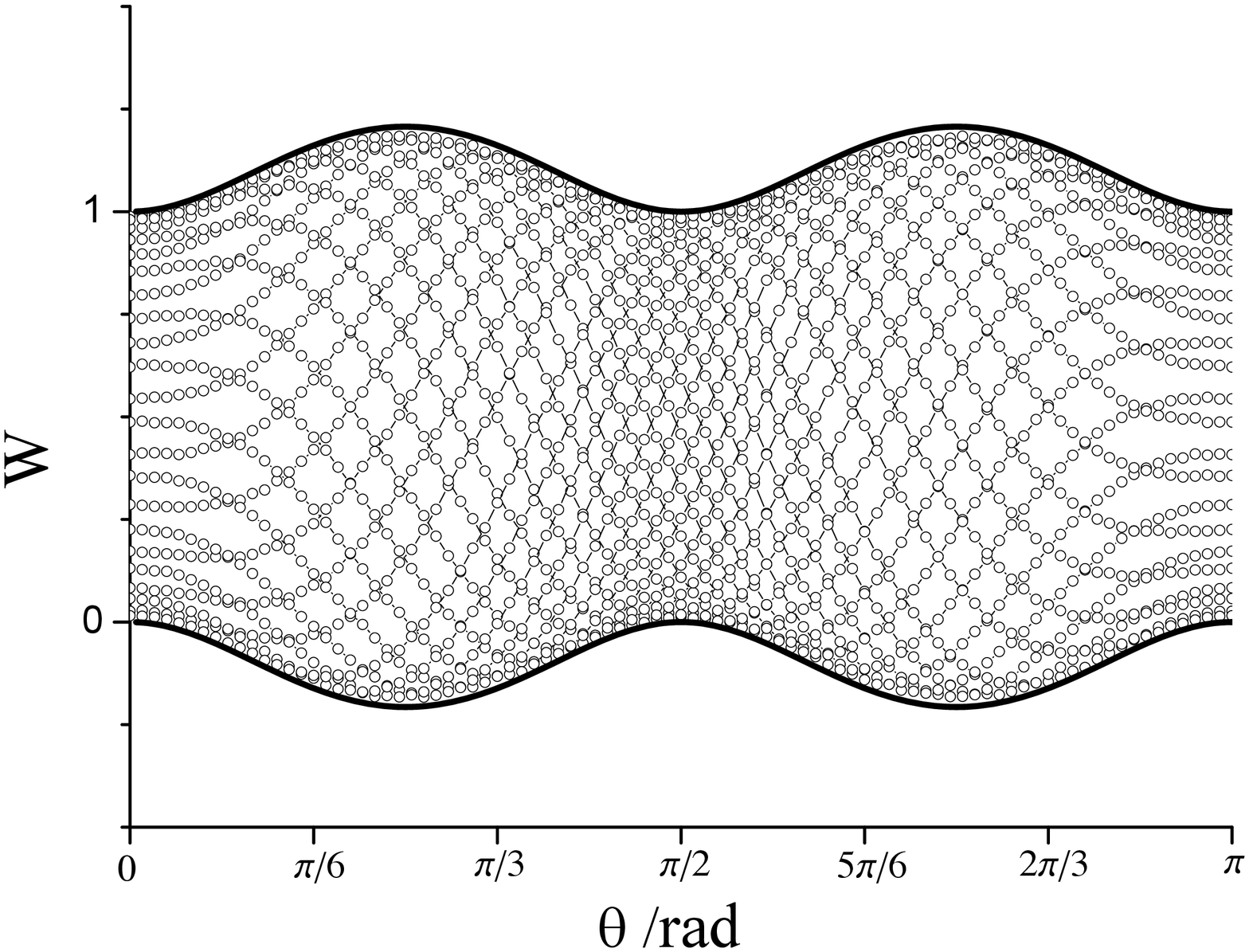}
\end{center}
\caption{ Experimental results. Each dot corresponds to an
experimental measurement of $_{A,B} \langle \varphi (\xi)
|\widehat{W}| \varphi (\xi) \rangle_{A,B}$ performed with a
different value of $\theta$ and $\xi$. Each curve corresponds to a
fixed value of $\xi$. Theoretical upper and lower $W$ bounds of
quantum correlation are also shown (thick black lines). }
\label{Figure 3}
\end{figure}

\begin{figure}[tbp]
\par
\begin{center}
\includegraphics[angle=0, width=10 cm]{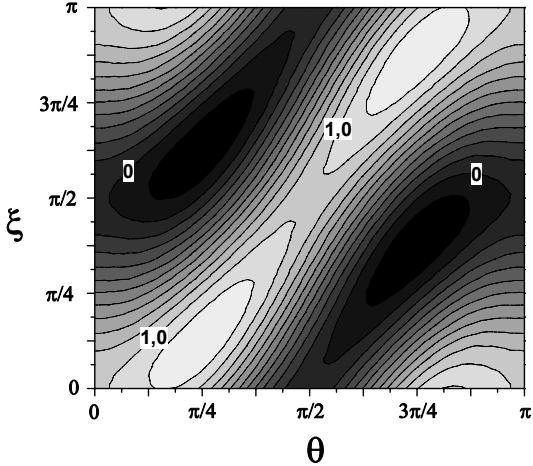}
\end{center}
\caption{ Contour plot obtained from the experimental data in Fig.
3. } \label{Figure 3}
\end{figure}

Let us now consider the problem of Ekert's protocol based QKD.
According to Ref.s \cite{EKE91, qk3, qk2, CDR03}, al least one of
the measurement settings of the Ekert's protocol should produce
binary non-local deterministic outputs, but local random outputs. In
general the state producing the maximal violation of the WI are not
suitable for QKD. Considering the four settings of the Wigner's
protocol $\mathcal{S}=(\{ -\theta_{A}, 0_{B}\}$, $\{ -\theta_{A},
\theta_{B}\}$, $\{ 0_{A}, 0_{B}\}$, $\{ 0_{A}, \theta_{B}\})$, the
states suitable for QKD should be of the form
\begin{eqnarray}
|\Gamma(\alpha_{A},\beta_{B},\gamma)\rangle_{A,B}&=&\frac{1}{\sqrt{2}}
( |s_{+}(\alpha)\rangle_{A}|s_{+}(\beta)\rangle_{B}  \nonumber \\
 & &+e^{i \gamma}|s_{-}(\alpha)\rangle_{A}|s_{-}(\beta)\rangle_{B} ),
\nonumber\\
|\Delta(\alpha_{A},\beta_{B},\delta)\rangle_{A,B}&=&\frac{1}{\sqrt{2}}
( |s_{+}(\alpha)\rangle_{A}|s_{-}(\beta)\rangle_{B} \nonumber \\
& &+e^{i \delta}|s_{-}(\alpha)\rangle_{A}|s_{+}(\beta)\rangle_{B} )
\label{QKDstate}
\end{eqnarray}
with $\{\alpha_{A}, \beta_{B}  \} ~ \in ~ \mathcal{S}$.

In the case of measurement settings $\{ -\theta_{A}, 0_{B}\}$, $\{
0_{A}, \theta_{B}\}$, for any value of $\theta$, $\gamma$ and
$\delta$, the states in Eq.  (\ref{QKDstate}) do not produce any
violation of WI, i.e. the states $|\Gamma( -\theta_{A},
0_{B},\gamma)\rangle_{A,B}$, $|\Delta( -\theta_{A},
0_{B},\delta)\rangle_{A,B}$ and $|\Gamma( 0_{A},
\theta_{B},\gamma)\rangle_{A,B}$, $|\Delta(0_{A},
\theta_{B},\delta)\rangle_{A,B}$ are not suitable to guarantee the
security of the Ekert's QKD protocol based on WI. By contrary
violations of WI are observed for settings $\{ 0_{A}, 0_{B}\}$, $\{
-\theta_{A}, \theta_{B}\}$.

In the case of setting $\{ 0_{A}, 0_{B}\}$ the maximal violation of
the lower bound achievable is $W=-0.125$, obtainable with the state
$|\Delta(0_{A},0_{B},\pi)\rangle_{A,B}$, corresponding to the
singlet state $| \psi^{-} \rangle_{A,B}$) when $\theta=\pi/6$ or
$5\pi/6 $ (lower thin black line in Fig. 1). Furthermore, at the
same angles the maximal violation of the upper bound achievable with
this setting ($W=1.125$) are predicted. The state producing these
violations is $|\Gamma(0_{A},0_{B},0)\rangle_{A,B}$, corresponding
to the triplet state $| \phi^{+} \rangle_{A,B}$ (upper thin black
line in Fig. 1).

Analogously,  for setting $\{ -\theta_{A}, \theta_{B}\}$ the maximal
violation of the lower bound ($W=-0.125$) is observed for the state
$|\Gamma(-\theta_{A}, \theta_{B},0)\rangle_{A,B}$ when
$\theta=\pi/3$ or $2\pi/3$ (lower dotted black line in Fig. 1),
while at the same angle the maximal violation of the upper bound
($W=1.125$) is predicted for the state $|\Delta(-\theta_{A},
\theta_{B},\pi)\rangle_{A,B}$ (upper dotted black line in Fig. 1).

This means that, in principle, there is not any advantage in using a
state different from the singlet state to implement the Ekert's QKD
protocol based on WI, or, equivalently, that the singlet state
naturally produced by type II PDC is one of the optimal state for
realizing QKD experiment with this protocol.

The last part of the paper is devoted to consider a more general
parametrization for WI. Let us consider the Wigner's operator
\begin{eqnarray}
\widehat{\mathcal{W}}&=&\widehat{P}_{+,A}(\alpha)\otimes
\widehat{P}_{+,B}(0)+\widehat{P}_{+,A}(0)\otimes
\widehat{P}_{+,B}(\beta) \nonumber \\& &+\widehat{P}_{-,A}(0)\otimes
\widehat{P}_{-,B}(0)- \widehat{P}_{+, A}(\alpha)\otimes
\widehat{P}_{+ ,B}(\beta), \label{WopG}
\end{eqnarray}
which is a generalization of the one in Eq. (\ref{Wop}).

As the Wigner's parameter consists of independent local projection
measurements on both Alice and Bob side, it's obvious that, by a
proper rotation of the Poincare's sphere at each side, any possible
choice of these projections measurements can be reduced to the ones
of the Eq. (\ref{WopG}). Thus, the most general parametrization of
the Wigner's parameter is equivalent to the one in Eq. (\ref{WopG}).

By the application of the $min/max$ principle, the bounds of quantum
correlation for $\mathcal{W}$ $(\mathcal{W}= \mathrm{Tr}[
\widehat{\mathcal{W}} ~ \widehat{\rho} ])$ are $-0.20711 \leq
\mathcal{W} \leq 1.20711$, as in the Filipp-Svozil parametrization.
This means that the generalization of the Wigner's parameter does
not provide any advantage in terms of increasing the region of
quantum correlation.

Eventually, also in the case of Ekert's QKD protocol based on WI
there is not any advantage in using the generalized version of the
Wigner's parameter with respect to the Filipp-Svozil one. In fact,
following the same line of thought developed in the case of the
Filipp-Svozil parametrization, the maximal violation for lower and
upper bound achievable by states suitable for QKD are $-0.125$ and
$1.125$, respectively.

In conclusion, we investigated theoretically and experimentally the
bounds of quantum correlation associated to the modified WI
according to Filipp-Svozil parametrization. We obtained the
configuration for the maximal violation of modified WI, which,
surprisingly, is not reachable by the singlet state as it happens
for CHSH inequality. Furthermore we extended our analysis to the
implementation of the Ekert's QKD protocol based on WI, and we
observed that, in this context, the singlet state allows the optimal
implementation of the QKD protocol. Furthermore we proved that there
is not any advantage in considering generalized parametrization with
respect to the Filipp-Svozil one.

\begin{acknowledgments}
We wish to thank M. Genovese for helpful discussion and to bring to
our attention Ref. \cite{Koc93}. Elsag Datamat carried out this work
within the contract Ministero della Difesa n. 9223 del 22.12.2005
ECFET QAP-2005-015848.
\end{acknowledgments}

\end{document}